\documentclass[10pt,twocolumn,letterpaper]{article}

\usepackage{cvpr}
\usepackage{times}
\usepackage{epsfig}
\usepackage{graphicx}
\usepackage{amsmath}
\usepackage{amssymb}

\usepackage{amsmath}
\usepackage{booktabs}
\usepackage[english]{babel}

\usepackage{xcolor}

\usepackage[pagebackref=true,breaklinks=true,letterpaper=true,colorlinks,bookmarks=false]{hyperref}

\cvprfinalcopy % *** Uncomment this line for the final submission

\def\cvprPaperID{2701} % *** Enter the CVPR Paper ID here

\ifcvprfinal\fi
\begin{document}

%%%%%%%%% TITLE
\title{Variational Autoencoders for Deforming 3D Mesh Models}

% \author{Qingyang Tan$^{*,\dagger}$
% 	, Lin Gao$^{*}$
% 	, Yu-Kun Lai$^{\ddagger}$, Shihong Xia$^{*}$ \\
% 	$^{\dagger}$ School of Computer and Control Engineering, University of Chinese Academy of Sciences \\
% 	tanqingyang14@mails.ucas.ac.cn \\
% 	$^{*}$ Beijing Key Laboratory of Mobile Computing and Pervasive Device, \\ 
% 	Institute of Computing Technology, Chinese Academy of Sciences \\
% 	\{gaolin, xsh\}@ict.ac.cn \\
% 	$^{\ddagger}$ School of Computer Science \& Informatics, Cardiff University \\
% 	Yukun.Lai@cs.cardiff.ac.uk
% }

\author{Qingyang Tan\textsuperscript{1,2}, Lin Gao\textsuperscript{1}\thanks{Corresponding Author}, Yu-Kun Lai \textsuperscript{3}, Shihong Xia\textsuperscript{1}$^{*}$\\
	\textsuperscript{1}Beijing Key Laboratory of Mobile Computing and Pervasive Device, \\
	Institute of Computing Technology, Chinese Academy of Sciences \\
	\textsuperscript{2}School of Computer and Control Engineering, University of Chinese Academy of Sciences\\
	\textsuperscript{3}School of Computer Science \& Informatics, Cardiff University\\
	tanqingyang14@mails.ucas.ac.cn, \{gaolin, xsh\}@ict.ac.cn, LaiY4@cardiff.ac.uk
}

\maketitle
%\thispagestyle{empty}

%%%%%%%%% ABSTRACT
\begin{abstract}
3D geometric contents are becoming increasingly popular. In this paper,  we study the problem of analyzing deforming 3D meshes using deep neural networks. Deforming 3D meshes are flexible to represent 3D animation sequences as well as collections of objects of the same category, allowing diverse shapes with large-scale non-linear deformations. We propose a novel framework which we call mesh variational autoencoders (mesh VAE), to explore the probabilistic latent space of 3D surfaces. The framework is easy to train, and requires very few training examples. We also propose an extended model which allows flexibly adjusting the significance of different latent variables by altering the prior distribution. Extensive experiments demonstrate that our general framework is able to learn a reasonable representation for a collection of deformable shapes, and produce competitive results for a variety of applications, including shape generation, shape interpolation, shape space embedding and shape exploration, outperforming state-of-the-art methods.
\end{abstract}

\section{Introduction}

With the development of performance capture techniques, 3D mesh sequences of deforming objects (such as human bodies) become increasingly popular, which are widely used in computer animation. Deforming meshes can also be used to represent a collection of objects
with  different shapes and poses, where  large-scale non-linear deformations are common. There are still challenging problems to analyze deforming 3D meshes  and synthesize plausible new 3D models.  From the analysis perspective, complex deformations mean it is difficult to embed such 3D meshes into a meaningful space using existing methods. Acquiring high quality 3D models is still time consuming, since multiple scans are usually required to address unavoidable occlusion. Thus effectively generating plausible new models is also highly demanded.
%And acquiring high quality 3D models is still time consuming, as due to unavoidable occlusion, scans from multiple views need to be fused to form a complete model.

%With the maturity of 3D acquisition techniques, 3D geometric contents are becoming increasingly popular. They are widely used in computer animation, manufacturing design, 3D printing etc.
%However, acquiring high quality 3D models is still time consuming, as due to unavoidable occlusion, scans from multiple views need to be fused to form a complete model. Effectively generating plausible new models is thus highly demanded.
In this paper, we propose a novel framework called mesh variational autoencoders (mesh VAE), which leverages the power of neural networks to explore the latent space behind deforming 3D shapes,
%shapes
 and is able to generate new models not existing in the original dataset. Our mesh VAE model is trained using a collection of 3D shapes with the same connectivity. Many existing deformable object collections satisfy this. They are not restricted to a single deformable object; examples include the MPI FAUST dataset \cite{Bogo:CVPR:2014} which includes human bodies of different shapes in different poses. In general, meshes with the same connectivity can be obtained using consistent remeshing.
Although convolutional neural networks (CNNs) have been widely used for image analysis and synthesis, applying them to 3D meshes is still quite limited.  Previous work focuses on generalizing CNNs from 2D to 3D while preserving image-like \emph{regular grid connectivity},
including 3D convolution over voxels (e.g.~\cite{wu20153d}), and 2D convolution over geometry images (e.g.~\cite{sinha2017surfnet}). However, the voxel representation is inefficient and given the practical limit can only represent rough 3D models without  details.
The parameterization process to generate geometry images involves unavoidable distortions and may destroy useful structure information.

Our work aims to produce a \emph{generative} model capable of analyzing model collections and synthesizing new shapes. To achieve this, instead of using representations with image-like connectivity, we propose to use a state-of-the-art surface representation called RIMD (Rotation Invariant Mesh Difference) \cite{Gao:2016:EFD:2965650.2908736} to effectively represent deformations, along with a variational autoencoder \cite{kingma2013auto}. To cope with meshes of arbitrary connectivity, we propose to use a fully-connected network, along with a simple reconstruction loss based on MSE (mean square error). As we will show later, the large number of coefficients in this network can be efficiently trained, even with a small number of training examples.  We also propose a new extended model where an additional adjustable parameter is introduced to control the variations of different latent variables in the prior distribution. This provides the flexibility of enforcing certain dimensions in the latent space to represent the most important differences in the original dataset.
As we will show later, by using an effective feature representation and a suitable network architecture, our general framework is able to produce competitive results for various applications, including shape generation, shape interpolation, shape embedding and shape exploration, outperforming state-of-the-art methods, where traditionally dedicated methods are needed for different applications.

\section{Related Work}\label{sec:rw}
\textbf{3D Shape Representation and Interpolation.} To effectively represent a collection of shapes with the same connectivity, a naive solution is to take the vertex positions as the representation. Such representations however are not translation or rotation invariant. Among various  representations, the RIMD representation \cite{Gao:2016:EFD:2965650.2908736} is translation and rotation invariant, and suitable for data-driven shape analysis. Therefore, we use it to represent shapes in our framework.

{A natural application for shape representations is to interpolate or blend shapes. Existing methods can be largely categorized into geometry-based (e.g.~\cite{huber2017smooth}) and data-driven (e.g.~\cite{CGF:CGF12991}) methods. The latter exploits latent knowledge of example shapes, and so can produce more convincing interpolation results when such examples are available. Our method is a generic framework and as we will show later, it produces comparable or even better results than state-of-the-art data-driven methods. Rustamov et al. \cite{Rustamov:2013:MEI:2461912.2461959} and Corman et al. \cite{Corman:2017:FCI:3068851.2999535} propose map-based methods to describe distortions between models, which can be used with PCA (Principal Component Analysis) to produce linear embedding, while our embedding application is non-linear.}

%Previous interpolations method including non data-driven ones like \cite{huber2017smooth}, or data-driven ones like \cite{CGF:CGF12991}. Latter methods could generate much more convincing interpolation results by analyzing the data set, so we choose them to compare with in the later section.}
% \GL{Recently, Huber et al. \cite{huber2017smooth} proposed a non data-driven approach to shape interpolation. Shape interpolation is only one application of our generic framework. Moreover, we have already compared our work with recent data-driven method (Gao et al. 2016a) in the paper, which is known to perform much better than non data-driven methods for challenging cases by exploiting knowledge.}

\textbf{Deep Learning for 3D Shape Analysis.} In recent years, effort has been made to generalize image-based CNN models to analyze 3D shapes. Su et al.~\cite{su2015multi} and  Qi et al.~\cite{qi2016volumetric} use multi-view CNN models for 3D object classification. Li et al.~\cite{li2015joint} produce joint embeddings of shapes and images, which are useful for shape-based image retrieval. Shi et al.~\cite{shi2015deeppano} propose a method that converts 3D shapes into a panoramic view, and propose a variant of CNN to learn  the representation from such views. Maturana and
Scherer~\cite{maturana2015voxnet} combine the volumetric occupancy grid representation with a 3D CNN to recognize 3D objects in real time. Wu et al.~\cite{wu2016single} interpret 3D object structure and 2D keypoint heatmaps from 2D images. Some other works generalize CNNs from the Euclidean domain to non-Euclidean domain \cite{boscaini2015learning,masci2015shapenet,boscaini2016anisotropic,boscaini2016learning,masci2015geodesic}, which is useful for 3D shape analysis such as establishing correspondences.
{Maron et al.~\cite{Maron:2017:CNN:3072959.3073616} define a natural convolution operator on surfaces  for CNNs by parameterizing the surface to a planar flat-torus. Our work has a rather different focus for analyzing and synthesizing 3D shapes with large deformations, although it may benefit from \cite{Maron:2017:CNN:3072959.3073616}  to generate consistent meshes.}
{Verma et al.~\cite{verma2017dynamic} propose convolutional filters which can be applied to nodes in a graph with different degrees so can be used in intrinsic geometry learning. However, their method is suitable for shape correspondence at the \emph{vertex} level (surface domain) rather than the \emph{shape} level (shape collection domain) and thus {cannot} be directly generalized to our problems such as shape generation and dimensionality reduction.}
{Tan et al.~\cite{DBLP:journals/corr/abs-1709-04304} propose a specific method to analyze sparse deformation patterns using CNNs, while our work provides a more generic framework for various applications.}

\textbf{3D Model Synthesis.} Traditional methods for 3D model synthesis use probabilistic inference. Xue et al.~\cite{xue2012example} recover 3D geometry from 2D line drawings using examples and a maximum-a-posteriori (MAP) probability optimization. Huang et al.~\cite{huang2015analysis} use a  probabilistic model to generate component-based 3D shapes. Such methods are generally suitable for specific types of 3D shapes.
Recent works consider exploiting the capability of deep learning for 3D model synthesis. For this purpose, the voxel representation is  resemblance to 2D images and widely used.  Wu et al.~\cite{wu20153d} propose a generative model based on a deep belief network trained on a large, unannotated database of voxelized 3D shapes. Girdhar et al.~\cite{Girdhar16b} jointly train an encoder for 2D images, and a decoder for voxelized 3D shapes, allowing 3D reconstruction from a 2D image. Yan et al.~\cite{NIPS2016_6206} generate 3D objects with deep networks using 2D images during training with a 3D to 2D projection layer. Rezende et al.~\cite{rezende2016unsupervised} learn strong deep generative models for 3D structures, which is able to recover 3D structures from 2D images via probabilistic inference. Choy et al.~\cite{choy20163d} develop a recurrent neural network which  takes in images of an object instance from arbitrary viewpoints and outputs a reconstruction of the object in the form of 3D occupancy grids. Sharma et al.~\cite{sharma2016vconv} propose a convolutional volumetric autoencoder without object labels, and learn the representation from noisy data by estimating the voxel occupancy grids, which is useful for shape completion and denoising. Such methods however suffer from high complexity of the voxel representation and can only practically synthesize coarse 3D shapes lack of details.

Sinha et al.~\cite{Sinha2016} propose to learn CNN models using geometry images~\cite{Gu02}, which is then extended to synthesize 3D models \cite{sinha2017surfnet}. Geometry images allow surface details to be well preserved. However, the parameterization used for generating geometry images involves unavoidable distortions and is not unique. This is more challenging for shapes with complex topology (e.g. high-genus models). Tulsiani et al.~\cite{tulsiani2016learning} develop a learning framework to abstract complex shapes  and assemble objects using 3D volumetric primitives.  Li et al.~\cite{li_sig17} propose a neural network architecture for encoding and synthesizing 3D shapes based on their structures. Nash et al.~\cite{nash2017shape} use a variational encoder to synthesize 3D objects segmented into parts. Both~\cite{li_sig17} and ~\cite{nash2017shape} are designed to synthesize man-made objects and require part-based segmentation as input, and cannot cope with unsegmented or deformed shapes addressed in this paper.
Such methods can produce shapes with complex structures, but the level of details is restricted to the components and primitives used.
We propose a general framework that  uses a rotation invariant mesh representation as features, along with a variational encoder to analyze shape collections and synthesize new models. Our method can generate novel plausible shapes with rich details, as well as a variety of other applications.

\section{Feature Representation}\label{sec:feature}
In this section, we briefly summarize the feature used for training neural networks, as well as reconstructing and generating new models.
We use the rotation-invariant mesh difference (RIMD) feature \cite{Gao:2016:EFD:2965650.2908736}.
We assume that we have $M$ ($M\geq 2$) models, with $n$ vertices that are in one-to-one correspondence. Generally, the first model is considered as the reference model and other models are deformed models. 
As we will show later, the choice of base mesh does not affect the results. The deformation gradient $\mathbf{T}_i$ at vertex $v_i$ (the $i$-th vertex) can be obtained by minimizing the energy:
\begin{equation}\label{eqn:energy1}
E(\textbf{T}_i) = \sum_{j\in N_i}c_{ij}\|\textbf{e}_{ij}' - \textbf{T}_i\textbf{e}_{ij}\|^2,
\end{equation}
where $N_i$ is the one-ring neighbors of vertex $v_i$, $\textbf{e}_{ij}' = \textbf{p}_i'-\textbf{p}_j'$, and $\textbf{e}_{ij} = \textbf{p}_i-\textbf{p}_j$. $\textbf{p}_i$ and $\textbf{p}_i'$ are the positions of $v_i$ on the reference and deformed models, respectively. $c_{ij}$ is the cotangent weight $c_{ij} = \cot\alpha_{ij} + \cot\beta_{ij}$, where $\alpha_{ij}$ and $\beta_{ij}$ are angles opposite to the edge connecting $v_i$ and $v_j$. {As shown in~\cite{Levi2015,Gao:2016:EFD:2965650.2908736}, the cotangent weight is helpful to avoid  discretization bias from underlying smooth surfaces to meshes.}
The deformation gradient can be decomposed into a rotation part and a scaling/shear part as $\textbf{T}_i = \textbf{R}_i\textbf{S}_i$, where the scaling/shear part $\textbf{S}_i$ is rotation invariant~\cite{Kelly2015}. The rotation difference  $dR_{ij}$ from $v_i$ to an adjacent vertex $v_j$ cancels out global rotation and is thus also rotation invariant:
$dR_{ij} = \textbf{R}_i^T\textbf{R}_j$.
The RIMD representation is obtained by collecting the {logarithm} of the rotation difference matrix $dR_{ij}$ of each edge ($v_i, v_j$) and scaling/shear matrix $\textbf{S}_i$ of each vertex $v_i$:
\begin{equation}\textbf{f} = \{\log dR_{ij};\textbf{S}_i\}\quad(\forall i, j\in N_i).\end{equation}
The use of matrix logarithm helps make the feature linearly combinable. Given a RIMD feature, which can be the output of our mesh VAE, the mesh model can be reconstructed efficiently by optimizing the energy in Eqn.~\ref{eqn:energy1}. See~\cite{Gao:2016:EFD:2965650.2908736} for details.
{As shown in Table 2 of~\cite{Gao:2016:EFD:2965650.2908736}, the mean reconstruction error between the reconstructed and ground truth shapes is extremely small ($10^{-4}$) with no visual difference, which shows the high representation power of RIMD feature.}

\section{Mesh VAE}\label{sec:meshvae}
In this section, we introduce our mesh VAE. We first discuss how to preprocess the feature representation to make it more suitable for neural networks, and then introduce the basic mesh VAE architecture, as well as extended models taking conditions into account  for improved control of the synthesized shapes and with improved embedding capability. Finally we summarize a variety of applications.

%, and talked about the modification we made to improve the embedding ability of the model.

\subsection{Feature Preprocessing}
In our architecture, we propose to use hyperbolic tangent ($tanh$) as the activation function of the probabilistic decoder's output layer. The range of $tanh$ is $(-1, 1)$, and it has the gradient vanishing problem during training when the output is near $-1$ and $1$. Although alternative activation function $ReLU$  does not have the gradient vanishing problem, its range of $(0, +\infty)$ is too large.
As a generative model, we expect the network output to be in a reasonable range.  With the limited range of the output, the broad range of $ReLU$ would make the model difficult to train.
The activation function $tanh$ is also applied in recent generation networks~\cite{pix2pix2016,li_sig17}.
%\GL{Alternative solutions are used, such as linear output layer for \cite{nash2017shape}, sigmoid layer for \cite{3dgan} and $tanh$ layer \cite{li_sig17}. }
To avoid the gradient vanishing problem of $tanh$, we use uniform normalization to preprocess the feature to the range of $[-a, a]$, where $a$ is close to but not equal to $1$. The preprocessing is a mapping as follows:
\begin{equation}
\widetilde{\textbf{f}_i^j} = 2a\times\frac{\textbf{f}^j_i - \min_j(\textbf{f}_i^j)}{\max_j(\textbf{f}_i^j)-\min_j(\textbf{f}_i^j)} - a,
\end{equation}
where $\textbf{f}^j_i$ is the $i$-th component of  model $j$'s RIMD feature, $\widetilde{\textbf{f}^j_i}$ is the preprocessed feature. If $\max_j(\textbf{f}_i^j) = \min_j(\textbf{f}_i^j)$ for  examples in the dataset, we replace them with $\textbf{f}_i^j \pm \epsilon$. Using $a$ and $\epsilon$ also allows the network to generate reasonable models with features outside the range of the original dataset. In our experiments, we choose $\epsilon = 10^{-6}$ and $a = 0.9$.

\begin{figure}[t]
\centering
\includegraphics[width = .35\textwidth]{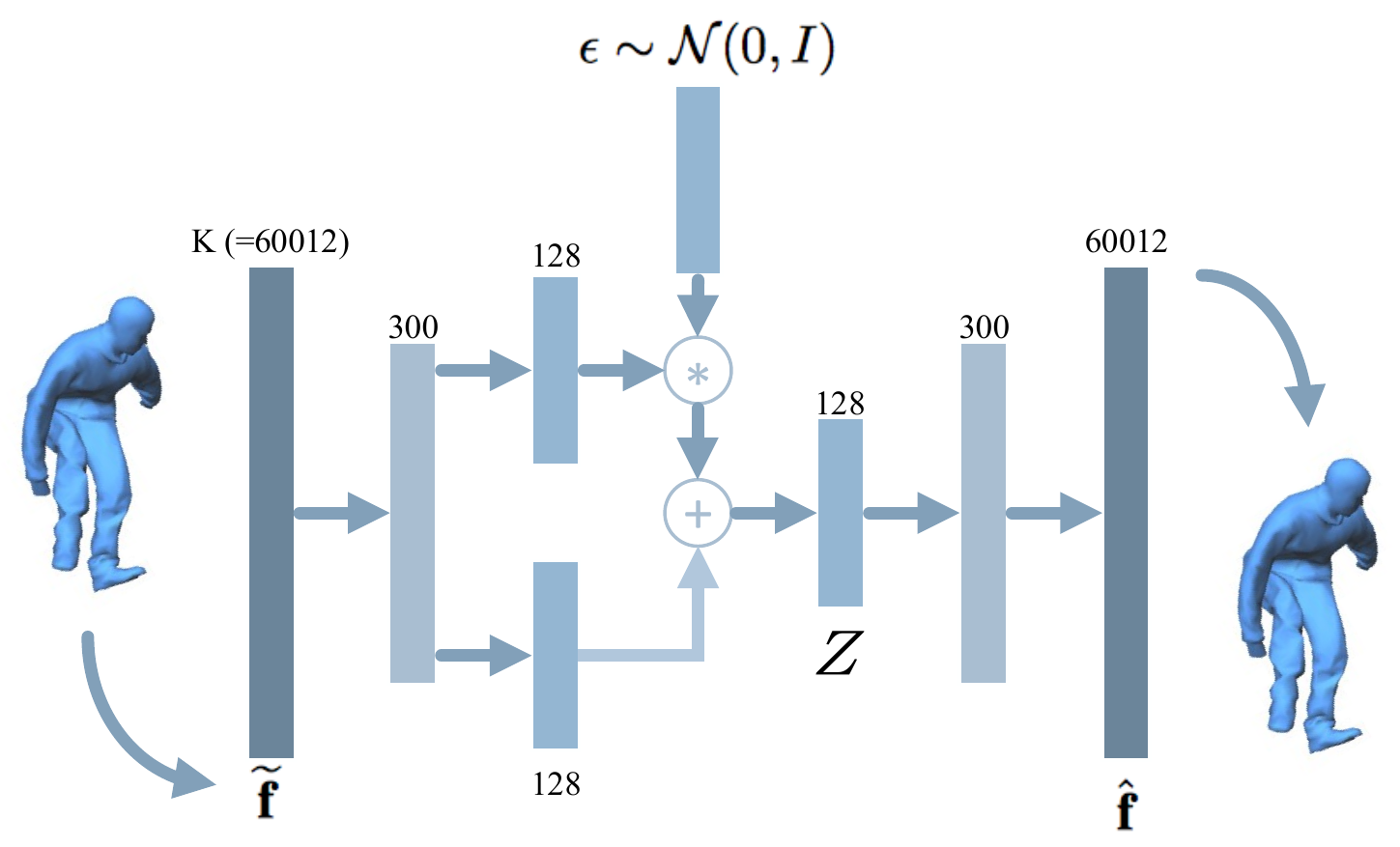}
\caption{Pipeline of mesh VAE. {The model in the figure has $2502$ vertices and $7500$ edges. Each $dR$ of a directed edge has $3$ elements, and $\textbf{S}$ of each vertex has $6$ elements, so the total feature dimension $K=2502\times 6+7500\times 3\times 2=60012$.}}\vspace{-2mm}
\label{pipeline}
\end{figure}

\subsection{Mesh VAE Architecture}
Following \cite{kingma2013auto}, our mesh VAE aims to find a probabilistic encoder and a probabilistic decoder. The encoder aims to map the posterior distribution from datapoint $x$ to the latent vector $z$, and the decoder produces a plausible corresponding datapoint $x$ from a latent vector $z$. In our mesh VAE, the datapoint $x$ is the preprocessed mesh feature $\tilde{\mathbf{f}}$. Denote by $K$ the dimension of $\tilde{\mathbf{f}}$. For mesh models with $5000$ faces, a typical input RIMD feature dimension is $K=60012$. By default,  we set the dimension of the latent space to $128$, which can be  adjusted proportionally with varying $K$.
In order to train a complicated model to fit the large dimension of mesh features, we replace the reconstruction loss from a probabilistic formulation to a simpler mean square error (MSE). The total loss function for the model is defined as
\begin{equation} \label{loss1}
L_{meshVAE} =
 \alpha\frac{1}{2MK}\sum_{j=1}^{M}\sum_{i=1}^{K}(\hat{\textbf{f}}_i^j - \widetilde{\textbf{f}_i^j})^2+D_{KL}(q(z|\widetilde{\textbf{f}})\|p(z)),
\end{equation}
where $\tilde{\textbf{f}}_i^j$ is the $i$-th dimension of model $j$'s preprocessed RIMD feature, $\hat{\textbf{f}}_i^j$ is the $i$-th dimension of model $j$'s output of the mesh VAE framework, $M$ is the number of models in the dataset, $\alpha$ is parameters to tune the priority over the reconstruction loss and latent loss,
$z$ is the latent vector,  $p(z)$ is the prior probability, $q(z|\widetilde{\textbf{f}})$ is the posterior probability, and $D_{KL}$ is the KL divergence. See \cite{kingma2013auto} for more details. The pipeline for our mesh VAE is summarized in Fig.~\ref{pipeline}.

\textbf{Network structure.} Since the RIMD features from different datasets have different neighboring structures, the spatial relationship inside the feature is not regular, so we use a fully-connected neural network for both the encoder and the decoder, expecting the network to learn suitable relationships,  rather than encoding such neighborhood structure in the network architecture explicitly.
%\GL{Similar architecture is adopted in \cite{nash2017shape} and \cite{li_sig17}.}
All the internal layers use batch normalization \cite{ioffe2015batch} and Leaky ReLU \cite{maas2013rectifier} as activation layers. The output of the encoder involves a mean vector and a deviation vector. The mean vector does not have an activation function, and  the deviation vector uses sigmoid as the activation function, which is then  multiplied by an upper bound value $\sigma_{max}$ of the expected deviation. The output of the decoder uses $tanh$ as the activation function, as we have discussed.

\textbf{Training details.}
For most tasks in this paper, we set $\alpha = 10^6${, which is optimized through unseen data to avoid overfitting}. We set the prior probability over latent variables to be $p(z) = \mathcal{N}(z;0,I)$ (Gaussian distribution with 0 mean and unit variation). We set $\sigma_{max} = 2$, such that  the output deviation can cover $1$, for  consistency with prior distribution. We set the learning rate to be $0.001$ and use ADAM algorithm \cite{Kingma2015} to train the model.

\subsection{Conditional Mesh VAE}
When synthesizing new shapes, it is often desirable to control the types of shapes to be generated, especially when the training dataset contains diverse shapes, e.g. the dataset from \cite{Dyna:SIGGRAPH:2015} which contains  different shapes and motion sequences.
%To allow users have more control over what kinds of models generated by the framework, especially for datasets have clear labels like s,
To achieve this, we extend our model to condition on labels in the training dataset following \cite{sohn2015learning}. Specifically, conditions are incorporated in both the encoder and the decoder as additional input, with the loss function  changed to
\begin{equation}
\begin{split}
L_{Conditional\ meshVAE} &=\\
\alpha\frac{1}{2MK}\sum_{j=1}^{M}\sum_{i=1}^{K}(\widehat{\textbf{f}_c}_i^j - &\widetilde{\textbf{f}_i^j})^2+D_{KL}(q(z|\widetilde{\textbf{f}}, c)\|p(z|c)),
\end{split}
\end{equation}
where $\widehat{\textbf{f}_c}$ is the output of the conditional mesh VAE, $p(z|c)$ and $q(z|\widetilde{\textbf{f}}, c)$ are conditional prior and posterior probabilities respectively.  %{, i.e. for a given condition $c$, $z$ is drawn from these two.} 
More experimental details are provided in Sec.~\ref{sec:exp}.

% \begin{figure}[t]
% \centering
% %\includegraphics[width = .35\textwidth]{evaluate.eps}
% \includegraphics[width = .25\textwidth]{evaluate.pdf}
% \caption{MSE reconstruction loss over training iterations.}
% \label{evaluate}
% \end{figure}
% \vspace{0pt}

\subsection{Extended Model with Improved Low-Dimensional Embedding}
The latent space $z$ provides an embedding that facilitates various applications. However, to ensure reconstruction quality, the dimension of $z$ is typically high (over 100), thus not suitable for low-dimensional embedding. To address this, we propose an extended model that modifies the prior distribution of the latent variables to  $p(z) = \mathcal{N}(z;0, diag(\sigma^2_{object}))$, where $\sigma_{object}$ is a tunable vector representing the deviation of the latent space, and $diag(\cdot)$ is the diagonal matrix of a given vector. Intuitively, if we set the $i$-th component ${\sigma_{object}}_i$ to be a small value, the approximate posterior $q(z_i|\widetilde{\mathbf{f}})$ will be encouraged to also have a small deviation, due to the latent cost.
As a result, in order to reduce the reconstruction loss, $z_i$ needs to be aligned to capture dominant changes. This behaviour is similar to PCA, although our model is non-linear and is thus able to better capture the characteristics of shapes in the collection.

\subsection{Applications}
%Our mesh VAE model has various applications, which are summarized below:
We summarize various applications of our mesh VAE.

\textbf{Generation and Interpolation.} Leveraging the capability of VAE, we can produce reasonable vectors in the latent space, and use them as input to the decoder to generate new models outside of the training dataset, which still satisfy the characteristics of the original dataset and are plausible. Meanwhile, similar to other autoencoder frameworks, we can interpolate the encoder results between two different shapes from the original dataset and then use the decoder to reconstruct  the interpolated mesh sequence. {In the same way as shapes in the dataset, our method can interpolate newly generated shapes by interpolating their latent vectors to produce a deformation sequence.}

\textbf{Embedding for Visualization and Exploration.}
Based on our extended model for low-dimensional embedding, we can map the distribution of shapes to a low-dimensional (typically two) latent space for visualization. We further allow users to easily browse the latent space to find the model they want, probably not in the original training dataset, using a synthesis based exploration approach. More details are provided in Sec.~\ref{sec:exp}.

% \begin{figure}[tb]
% \centering
% \includegraphics[width = .25\textwidth]{prcurve.pdf}
% \caption{Precision Recall Curve with style and pose as labels using 2D embedding of Dyna dataset.}
% \label{embpr}
% \end{figure}
% \vspace{0pt}

% \begin{table}[t]
% \centering
% \begin{tabular}{lcccc}
% \toprule
%  &\multicolumn{4}{c}{$\alpha$} \\
% \cmidrule(r){2-5}
% Dataset    & $10^8$ & $10^7$ & $10^6$ & $10^5$\\
% \midrule
% Jumping \cite{Vlasic2008} & $1.1385$ & $1.1319$ & $1.1090$ & $1.1884$ \\
% \bottomrule
% \end{tabular}
% \caption{Reconstruction error on held-out shapes with different $\alpha$ in Eq. \ref{loss1}.}
% \label{alpha_value}
% \end{table}

\begin{table}[t]
\centering
\begin{tabular}{lccc}
\toprule
 &\multicolumn{3}{c}{Latent Dimension} \\
\cmidrule(r){2-4}
Dataset    & $16$ & $128$ & $256$\\
\midrule
Jumping \cite{Vlasic2008} ($\times10^{-4}$) & $4.9129$ & $4.4325$ & $4.2442$\\
Face \cite{zhang-siggraph2004-stfaces} ($\times10^{-2}$) & $8.3833$ & $7.8025$ & $7.9192$\\
\bottomrule
\end{tabular}
\caption{Per-vertex reconstruction error on held-out shapes with different latent dimensions.}\vspace{-2mm}
\label{latent_dim}
\end{table}

\begin{table}[t]
\centering
\begin{tabular}{lccc}
\toprule
 &\multicolumn{3}{c}{Feature} \\
\cmidrule(r){2-4}
Dataset {[\#. Shapes]} & \footnotesize{3D}  & \footnotesize{Aligned}  & \small{RIMD} \\
 & \footnotesize{Coordinates} & \footnotesize{Coordinates} & \\
\midrule
\footnotesize{SCAPE [$71$] \cite{SCAPE}($\times10^{-3}$)} & \small{$15.7249$} & \small{$7.3582$} & \small{$\textbf{3.7418}$} \\
\footnotesize{Jumping [$150$] \cite{Vlasic2008} ($\times10^{-4}$)} & \small{$49.4876$} & \small{$11.2226$} & \small{$\textbf{4.4325}$}\\
\footnotesize{Bouncing [$75$] \cite{Vlasic2008} ($\times10^{-4}$)} & \small{$28.1823$} & \small{$6.6970$} & \small{$\textbf{3.7358}$}\\
\footnotesize{Face [$385$] \cite{zhang-siggraph2004-stfaces} ($\times10^{-2}$)} & \small{$47.1863$} & \small{$27.8445$} & \small{$\textbf{7.8025}$}\\
\footnotesize{Flag [$90$] ($\times10^{-3}$)} & \small{$29.5490$} & \small{$2.4462$}  & \small{$\textbf{1.7916}$}\\
\footnotesize{Dyna [$2168$] \cite{Dyna:SIGGRAPH:2015} ($\times10^{-3}$)} & \small{$2.9488$} & \small{$1.7634$}  & \small{$\textbf{1.2128}$}\\
\bottomrule
\end{tabular}
\caption{Per-vertex reconstruction error on held-out shapes with different feature representations.}
\label{feature_evaluation}
\end{table}

\begin{table}[t]
\centering
\begin{tabular}{lccccc}
\toprule
Base Mesh &\multicolumn{3}{c}{No. $1$} & No. $6$ & No. $72$  \\
Points & 6890 & 6002 & 8002 & 6890 & 6890\\
\midrule
Error($\times 10^{-4}$) & $2.77$ & $2.49$ & $3.01$ & $2.85$ & $2.80$\\
\bottomrule
\end{tabular}
\caption{{Per-vertex reconstruction error on held-out shapes of chicken wings dataset from \cite{Dyna:SIGGRAPH:2015} with different mesh densities or base mesh choices.}}
\label{base_mesh}\end{table}

\section{Experiments}\label{sec:exp}

\subsection{Framework Evaluation}\label{sec:evaluation}
To more thoroughly justify the  design choices of our framework and demonstrate the effectiveness of  using RIMD features, we compare results with different parameters, settings and input within our framework.

\textbf{Latent Dimensions.} We compare the reconstruction loss (per-vertex position errors) on held-out models with different latent dimensions (see Table~\ref{latent_dim}). This suggests that using 128 dimensions is effective in improving reconstruction quality. Lower dimensions cannot capture enough information, while higher dimensions cannot improve results much and can cause overfitting.

\textbf{MSE Loss.} We compare MSE loss we use with alternative probabilistic reconstruction loss following \cite{gregor2015draw} which was originally used for images. The RIMD features are mapped to $[0.1, 0.9]$ similar to probability, with sigmoid as output layer, and cross entropy as reconstruction loss. However, this method does not converge even with three times of epochs we use for MSE loss.
\begin{figure}[tb]
\centering
\includegraphics[width = 0.36\textwidth]{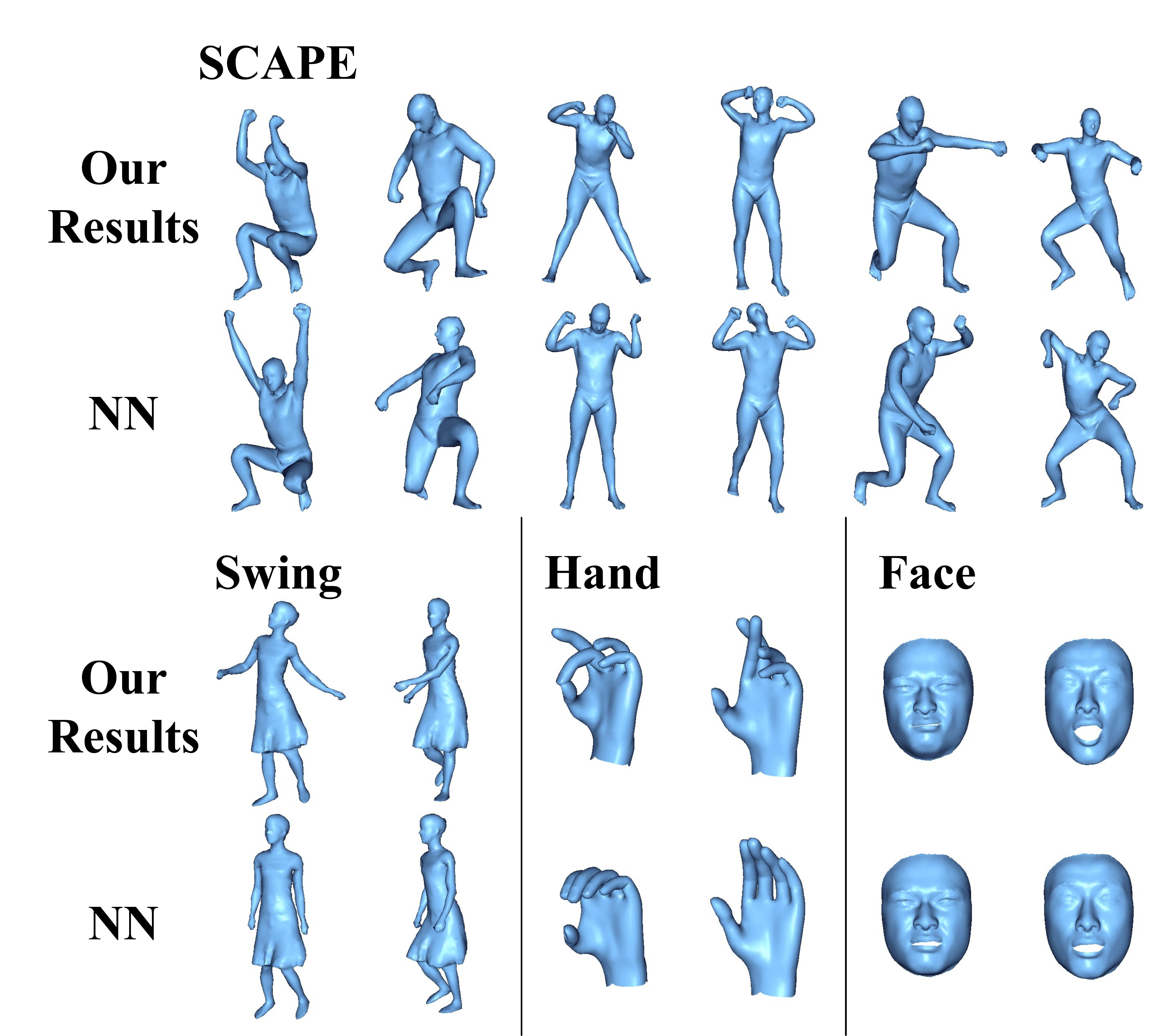}
\caption{Randomly generated new shapes using our framework, with their nearest neighbors in the original dataset shown below.}\vspace{-2mm}
\label{random}
\end{figure}
\vspace{0pt}

\begin{figure}[tb]
\centering
\includegraphics[width = 0.4\textwidth]{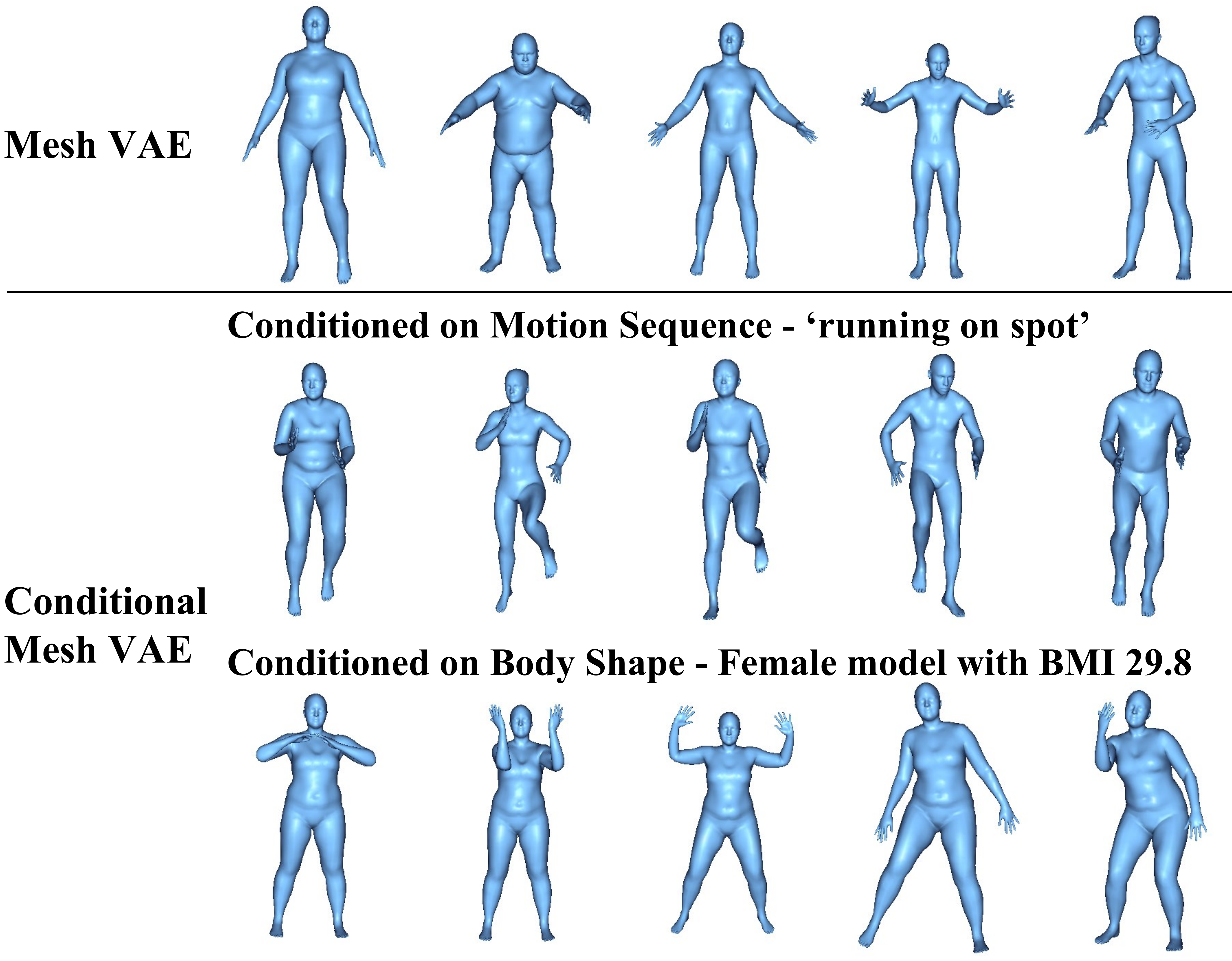}
\caption{{Randomly generated new shapes using mesh VAE and conditional mesh VAE.}}
\label{random_con}
\vspace{-2mm}
\end{figure}

% \paragraph{Balancing losses.}\GL{$\alpha$ in Eq. \ref{loss1} is the parameter to balance reconstruction and latent losses. We use unseen data to optimize $\alpha$ without causing overfitting. As the results in Table 2, $\alpha = 10^6$ shows the best result for the Jumping from \cite{Vlasic2008}.}
\begin{figure*}[tb]
\centering
\includegraphics[width = 0.75\textwidth]{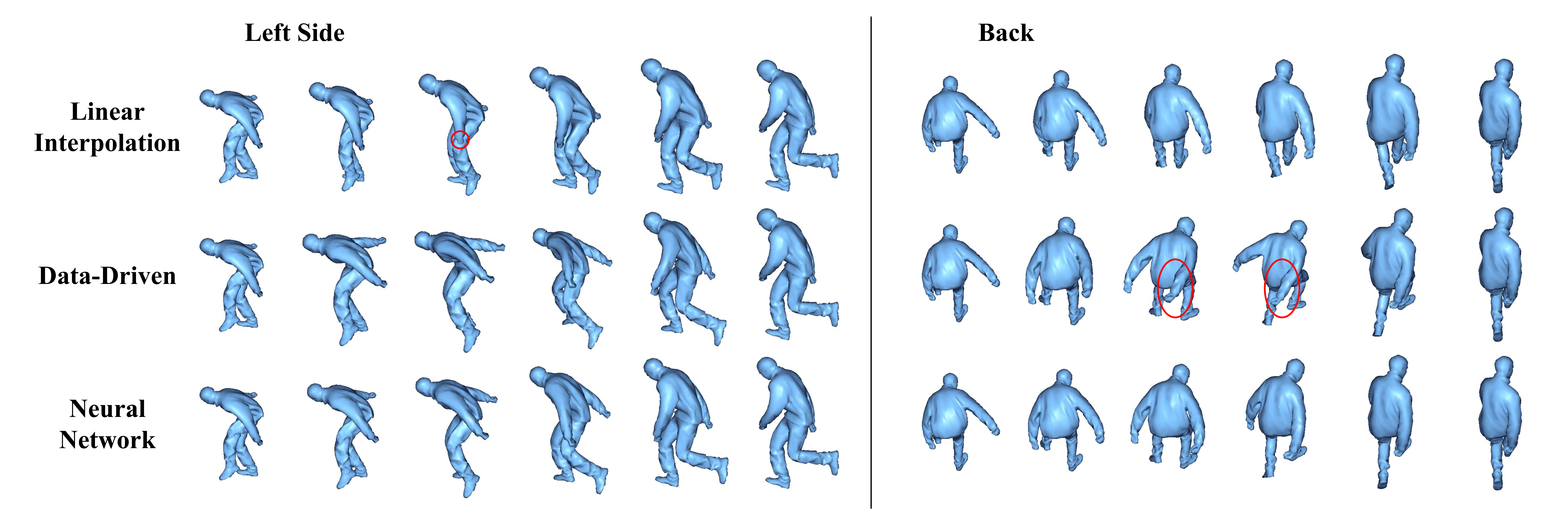}
\caption{Comparison of mesh interpolation results. First row is the result of direct linear interpolation of the RIMD feature, second row is the result of \cite{CGF:CGF12991}, third row is our result. All the results are shown in two views, with artifacts marked using red circles.}
\label{inter}\vspace{-2mm}
\end{figure*}

\begin{figure}[tb]
\centering
\includegraphics[width = .35\textwidth]{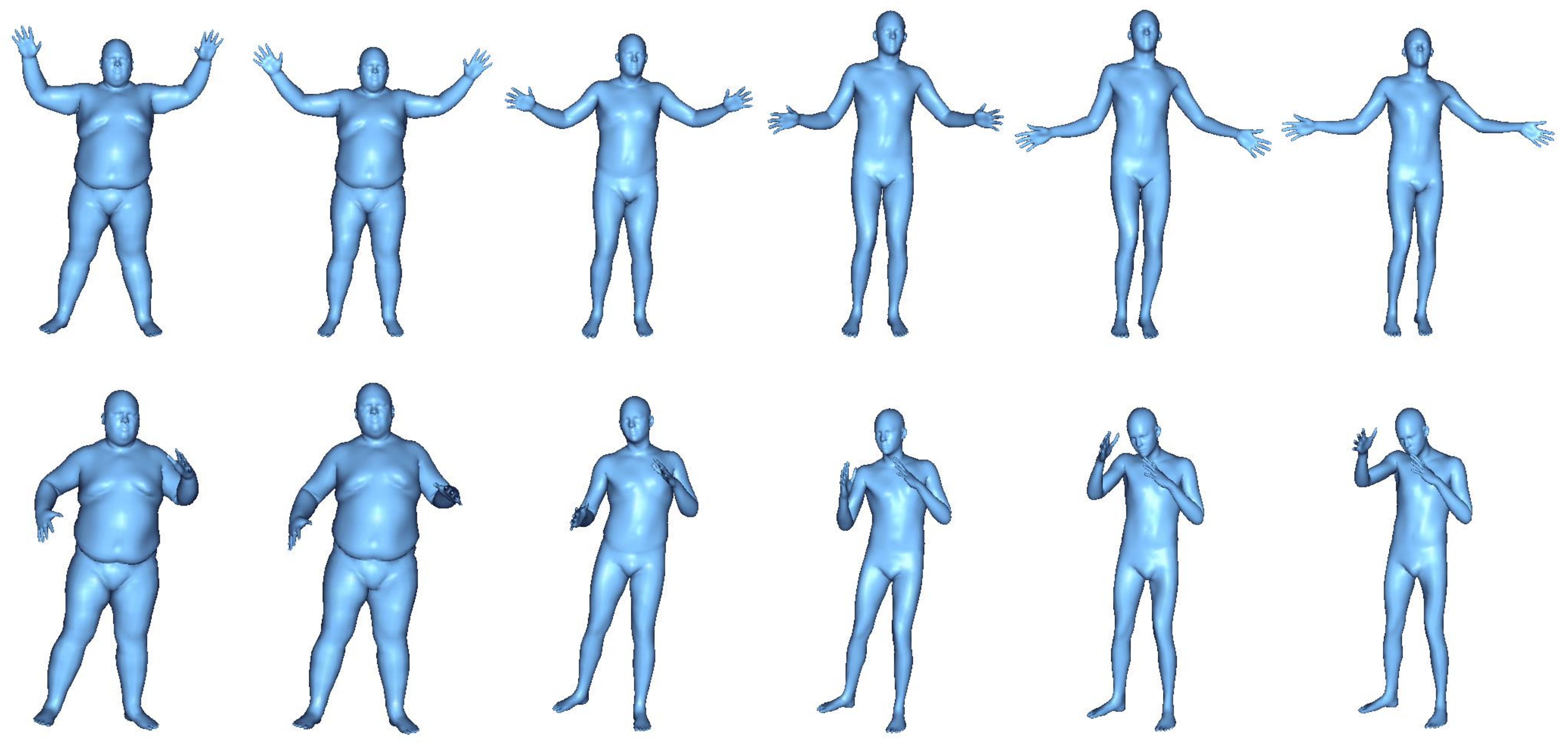}
\caption{Inter-class interpolation with different body shapes and poses.}
\label{interclass}\vspace{-2mm}
\end{figure}

\begin{figure}[tb]
\centering
\includegraphics[width = .4\textwidth]{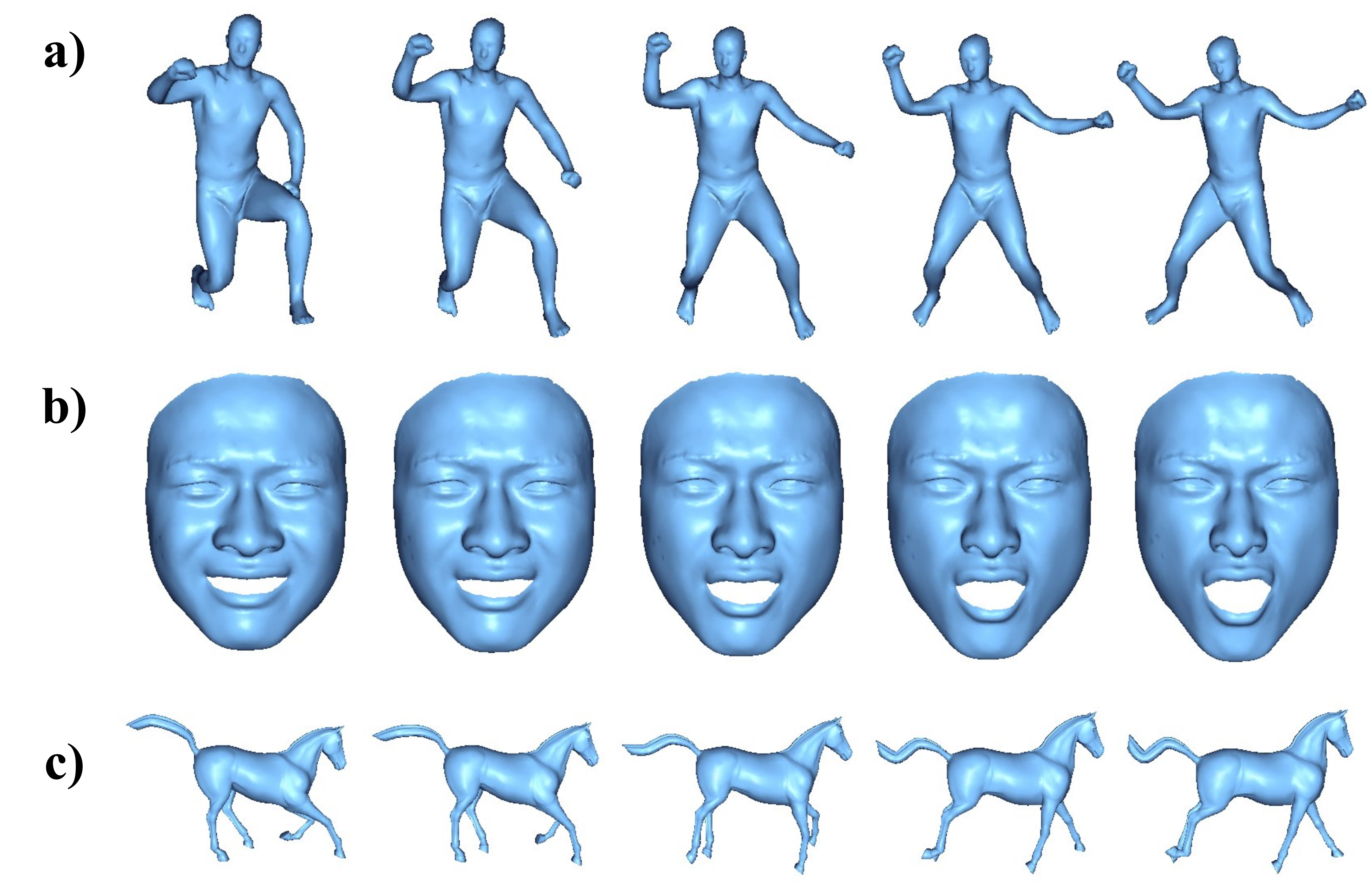}
\caption{{More interpolation examples. (a) Results between newly generated shapes. (b-c) Models other than human bodies.}}
\label{more}\vspace{-3mm}
\end{figure}
\textbf{RIMD Feature.} To verify the effectiveness of the RIMD feature, we work out per-vertex position errors of held-out shapes in different datasets. We compare our method using RIMD feature with baseline methods using 3D vertex coordinates and aligned coordinates (by finding global rotations that minimize vertex differences) as input,  with network layers and parameters adjusted accordingly to optimize their performance. In all the datasets we tested, our method produces visually high quality meshes, substantially reducing reconstruction loss by $59\%-94\%$ compared with 3D coordinates and $27\%-72\%$ compared with aligned coordinates. The details are shown in Table~\ref{feature_evaluation}. {Our proposed framework is easy to train, as it can be effectively trained even with small datasets as these examples in the paper typically contain $70$-$150$ models, while avoiding overfitting as demonstrated by the generalizability.} %{We have to point out that the size of the dataset is largely restricted by the available datasets. We also tested the generalizability on larger Dyna ~\cite{verma2017dynamic} dataset with different motions of different male body shapes with results mentioned in last row of Table ~\ref{feature_evaluation}, which proves that our method remains robust with increasing complexity.} 
The dataset size is largely restricted by data availability. Our method has good generalizability also for larger Dyna dataset~\cite{verma2017dynamic} with different motions of different male body shapes (see last row of Table ~\ref{feature_evaluation}).
{We test our method on simplified, original and refined chicken wings dataset from~\cite{Dyna:SIGGRAPH:2015} with $6002$, $6890$ and $8002$ points, and also compare the results when different base meshes are chosen, the per-vertex reconstruction errors of unseen data are shown in Table~\ref{base_mesh}. The differences of per-vertex errors between different density or base choices are below $10^{-4}$, with no visual difference. 
%This demonstrates that our results do not depend on the density of meshes or the choice of the base mesh.
This demonstrates that our method is robust with such changes.
}

% \subsection{Generalization Ability}
% To test the generalization ability of the network, we divide a given shape collection into training and validation sets, and reconstruct shapes not seen during the training. We use the chicken wings dataset from~\cite{Dyna:SIGGRAPH:2015}, with $172$ training models and $44$ validation models. The MSE reconstruction loss {of the RIMD feature} over training iterations is shown in Fig.~\ref{evaluate}. {Our framework is efficient to train. The whole training time is only 36.98 minutes using a computer with an Intel Xeon E5-2620 CPU and an NVIDIA Tesla K40C GPU.} We can see that the network has a fairly strong generalization ability as the loss for unseen shapes is reasonably low. Note that the total loss involves both the reconstruction loss and the latent loss, so the reconstruction loss may sometimes fluctuate during optimization.

\subsection{Generation of Novel Models}
\textbf{Mesh VAE.} We use standard setting $z\sim \mathcal{N}(0,I)$ as the input to the probabilistic decoder, and test the capability of the mesh VAE framework to generate new models. We train the network on the SCAPE dataset~\cite{SCAPE}, `Swing' dataset from~\cite{Vlasic2008}, face dataset~\cite{zhang-siggraph2004-stfaces} and hand dataset. The results are shown in Fig.~\ref{random}. We can see that, since we learn the latent space of 3D models from our proposed framework, we can easily generate new models, which are generally plausible. To validate that  the network does not only memorize the dataset, we also show the models with their nearest neighbors in the original dataset. The nearest neighbors are based on the Euclidean distance in the RIMD feature space, following \cite{Gao:2016:EFD:2965650.2908736} which considers this as a suitable measure to evaluate results related to RIMD features. The model in the original dataset with the shortest distance to the newly generated model is chosen as the nearest neighbor. It is clear that mesh VAE can generate plausible new models by combining local deformations of different model parts.
\begin{figure*}[tb]
\centering
\includegraphics[width = .7\textwidth]{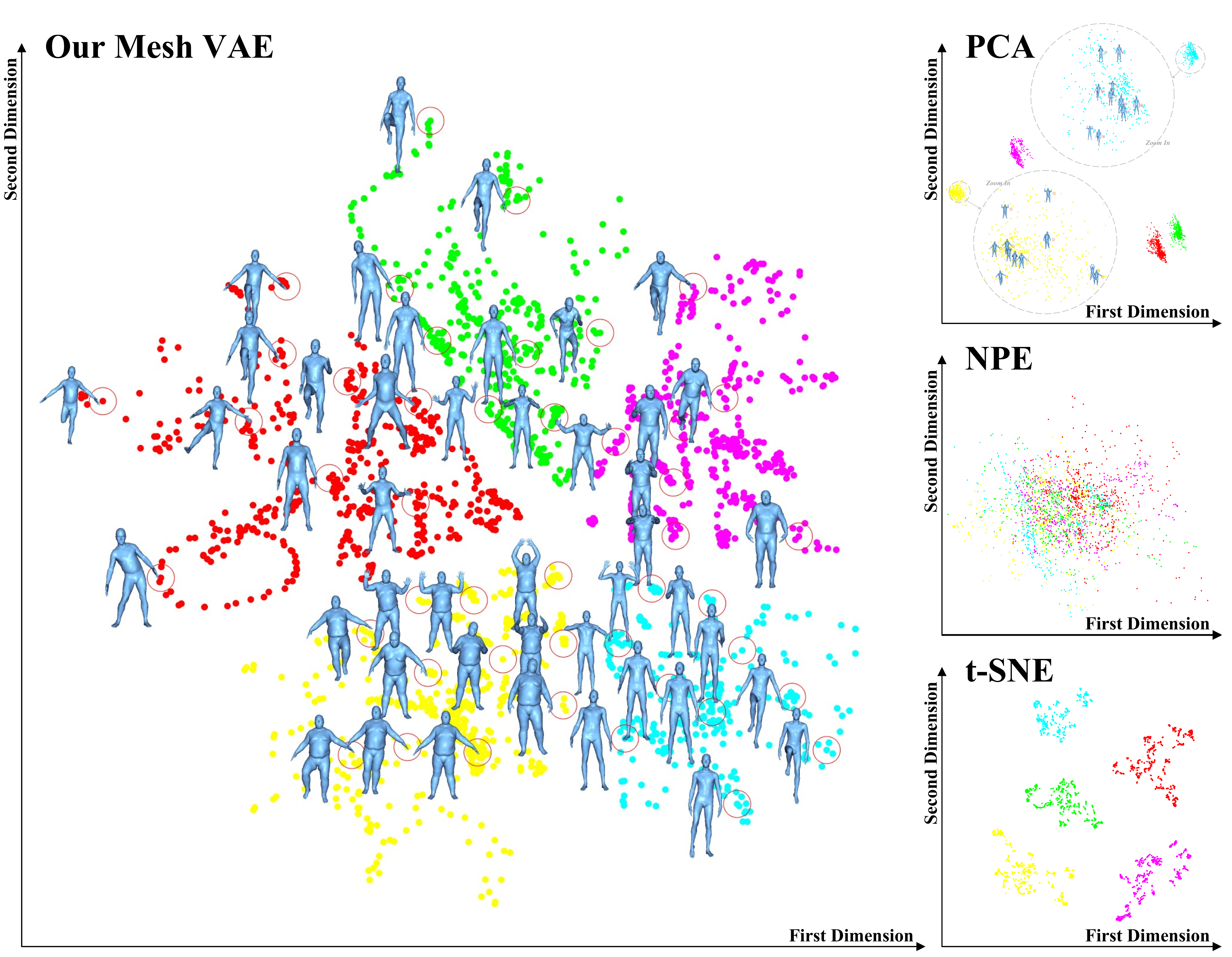}
\caption{2D embedding of Dyna dataset. Different colors represent different body shapes. PCA embedding is very sparse so we include  two zoom-in subfigures as shown in the circles.}\vspace{-3mm}
\label{embfig}
\end{figure*}
\vspace{0pt}

\textbf{Conditional Mesh VAE.} With the help of labels as conditions, users can generate more specific models. To show this, we train the conditional mesh VAE on the Dyna dataset from~\cite{Dyna:SIGGRAPH:2015},{ which contains male/female models with different BMI performing different motions. All the models share the same connectivity. We used BMI+gender and motion as conditions to train a network for all models at the same time. We then randomly generate models either conditioned on action with the label  `running on spot' which contains shapes of a running action, as well as on body shape `50022' --- a female model with BMI $29.8$, and compare the generation results from the normal mesh VAE. As the results in Fig.~\ref{random_con} show, more specific models can be synthesized with the given conditions.}

% conditioned on shapes and motion sequences. We then randomly generate models either conditioned on action with the label  `running on spot' which contains shapes of a running action, as well as  on body shape `50022' --- a female model with BMI $29.8$, and compare the generation results from the normal mesh VAE. As the results in Fig \ref{random_con} show, more specific models can be synthesized with the given conditions.

\subsection{Mesh Interpolation}
We test the capability of our framework to interpolate two different models in the original dataset. First, we generate the mean output of the probabilistic encoder for $2$ models in the original dataset. Then, we linearly interpolate between the two outputs and generate a list of inputs to the probabilistic decoder. Finally, we use the outputs of the probabilistic decoder to reconstruct a 3D deformation sequence. We compare the results on `jumping' dataset from~\cite{Vlasic2008} with direct linear interpolation of the RIMD feature, and the state-of-the-art data-driven deformation method~\cite{CGF:CGF12991}, as shown  in Fig.~\ref{inter}. We can see that direct linear interpolation produces models with self-intersections. {Traditional geometry-based methods such as as-rigid-as-possible interpolation~\cite{Liu2011} have similar problems (results omitted due to space restriction).} The interpolation result in the latent space can avoid these problems. Meanwhile, the data-driven method tends to follow the movement sequences from the original dataset which has similar start and end states, and the results have some redundant motions such as the swing of right arm. Our interpolation result gives  a reasonable motion sequence from start to end.
We also test inter-class interpolation on `jumping jacks' and `punching' datasets from \cite{Dyna:SIGGRAPH:2015}, interpolating between models with different body shapes and poses. The results are shown in Fig.~\ref{interclass}.
{We show more interpolation results in Fig.~\ref{more}, including sequences between newly generated models and models beyond human bodies.}

\begin{figure*}[tb]
\centering
\includegraphics[width = .7\textwidth]{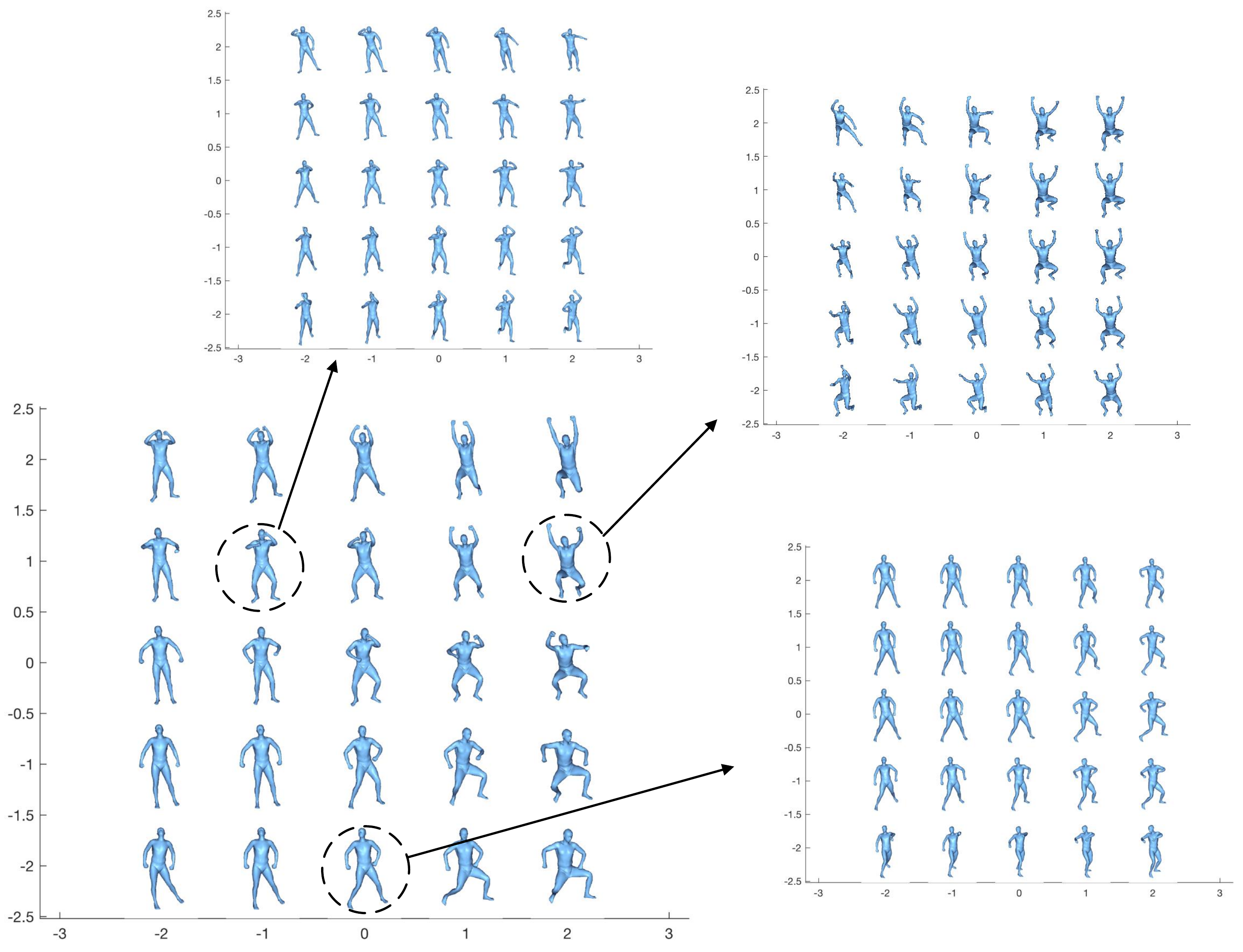}
\caption{Exploring the latent shape space. {The main figure (bottom left) shows the exploration results of the first two dimensions, and the subfigures are results of browsing the third and fourth dimensions, with the first two dimensions fixed based on the selected shape.}}
\label{browse1}\vspace{-2mm}
\end{figure*}
\vspace{0pt}
\subsection{Embedding}
As mentioned in Sec.~\ref{sec:meshvae}, we adjust the deviation of latent probability to let some dimensions of $z$  capture more important deformations. We utilize this capability to embed shapes in a low-dimensional space. We use different motion sequences of different body shapes from \cite{Dyna:SIGGRAPH:2015}, and compare the results of our method with standard dimensionality reduction methods PCA, NPE (Neighborhood Preserving Embedding) {and t-SNE (t-Distributed Stochastic Neighbor Embedding)}. When training meshVAE, we set the deviation of the first two dimensions of the latent space $z$ to be $0.1$, and the remaining $1$. {The K-nearest neighbor setting for NPE is set to $k=25$ and the perplexity for t-SNE is set to 30, which are optimized to give more plausible results.} The comparative results are shown in Fig.~\ref{embfig}. Since the dataset contains a large number of diverse models, our embedding is able to capture more than one major deformation patterns in one dimension. When using the first two dimensions for visualization, our method effectively divides all the models according to their shapes, while allowing models of similar poses to stay in close places. PCA is a linear model, and can only represent one deformation pattern in each direction. In this case, it can discriminate body shapes but the distribution is rather sparse. It also cannot capture other modes of variation such as poses in 2D embedding. {The results of t-SNE have relatively large distances between primary clusters, similar to PCA. This phenomenon is also mentioned in~\cite{wattenberg2016how}.} NPE cannot even discriminate body shapes, which shows the difficulty of the task. To quantitatively analyze the performance of embeddings, we perform experiments on the retrieval task using styles and poses as labels. {The AUC (area under the curve) values of the precision-recall curves are: our method (0.5168) $>$ t-SNE (0.4961) $>$ PCA (0.2272) $>$ NPE (0.1391).}
%The precision-recall curves are shown in Fig.~\ref{embpr}, which shows our embedding is more discriminative than alternative methods.

\subsection{Synthesis based Exploration of the Shape Space}
By adjusting the parameters $\sigma_{object}$, our meshVAE model is able to capture different importance levels of features, and thus allows the user to easily explore the latent space of the dataset, and find new models they want. We test this exploration application over the SCAPE dataset \cite{SCAPE} and demonstrate the results in 2 levels (each in a 2D space). We set the first two dimensions of $\sigma_{object}$ to $0.1$, the next two dimensions $0.5$, and remaining  $1$. In the main figure of Fig.~\ref{browse1}, we show the result of browsing the latent space in the first two dimensions. We set the range for both dimensions as $[-2, 2]$, while ignoring the remaining dimensions. We can see that these feature dimensions correspond to dominant shape deformation. The first dimension appears to control the total height of the model; when the value increases, the model changes from standing to squatting. The second dimension appears to control the supporting leg; when the value increases, the model starts to change its supporting leg from left to right. After picking up interested locations in the first level browsing, the user can fix the values of the first two dimensions, and use the next two dimensions to explore the shape space in more detail. In the subfigures of Fig.~\ref{browse1}, we show the results of browsing the third and fourth dimensions of the latent space. The values of the first two dimensions are selected based on the choice in the first step, and then the range for the third and fourth dimensions is set to $[-2,2]$. The third dimension appears to control the height of arms; when the value increases, the model gradually lifts the arms.  The fourth dimension appears to control the direction of the body; when the value increases, the model gradually turns to the left.

\section{Conclusions}\label{sec:conc}
In this paper, we introduce mesh variational autoencoders (mesh VAE), which use a variational autoencoder model with a mesh-based rotation invariant feature representation. We further study an extended model where the variation of latent variables can be controlled. We demonstrate that our generic model has various interesting applications, including analysis of shape collections as well as generating novel shapes. Unlike existing methods, our method can generate high quality deformable models with rich details. Experiments show that our method outperforms state-of-the-art methods in various applications. A restriction of our mesh VAE model is that it can only process homogeneous meshes.  As future work, it is desirable to develop a framework capable of handling shapes with different topologies as input.

\section*{Acknowledgments}
This work was supported by grants from the National Natural Science Foundation of China (No. 61502453, No. 61772499 and No.61611130215), Royal Society-Newton Mobility Grant (No. IE150731), the Science and Technology Service Network Initiative of Chinese Academy of Sciences (No. KFJSTS-ZDTP-017).

{\small
\bibliographystyle{ieee}
\bibliography{meshVAE}
}

\end{document}